\documentclass[prd,reprint,showpacs,showkeys]{revtex4}
\usepackage{graphics}
\usepackage{eurosym}
\usepackage{amsfonts}
\usepackage{mdframed}
\usepackage{amssymb}
\usepackage{amsmath}
\usepackage{graphicx}
\usepackage[font={footnotesize,it}]{caption}

\begin{document}

\title{Entangled Particles Tunneling From a Schwarzschild Black Hole
immersed in an Electromagnetic Universe with GUP }
\author{A. \"{O}vg\"{u}n}
\email{ali.ovgun@emu.edu.tr}
\affiliation{Physics Department, Eastern Mediterranean University, Famagusta, Northern
Cyprus, Mersin 10, Turkey.}
\date{\today }

\begin{abstract}
Quantum gravity has exciting peculiarities on the Planck scale.The effect of
generalized uncertainty principle (GUP) to the entangled scalar/fermion particles'
tunneling from a Schwarzschild black hole immersed in an electromagnetic
Universe is investigated by the help of semi-classical tunneling method. The
quantum corrected Hawking temperature of this black hole with an external
parameter "a" modifies the Hawking temperature for the entangled
particles.
\end{abstract}



\maketitle

\section{Introduction}

A breathtaking process like white rabbit and black magic, first theorized by
Stephen Hawking, by which a black hole can emit particles \cite{hw1}.
Hawking originally have used Bogoliubov's method \cite{hw2}, however, after
that several methods of deriving Hawking radiation appeared. Understanding
the Hawking radiation is a subject of long interest and the tunneling
phenomenon has been extensively studied and it is applied on various black
holes and also wormholes \cite%
{DR,pw,sri,sh1,s1,Mazhari,s3,jiang,s5,angh1,pad2,mann1,mann2,kr1,kr2,a5,a2,s6,s7}%
. Today, one of the main challenges in physics is to merge quantum theory
and the theory of general relativity into a unified framework, which should
be modified with a minimum length scale of the order of the Planck length.
So that the minimal length is developed on a very strong background as a
quantum gravity such as string theory, loop quantum gravity and a
non-commutative geometry \cite{gidd1,rov}. This length derived from some
different ways such as a generalized uncertainty principle (GUP), an
extended uncertainty principle (EUP) and a generalized EUP (GEUP) \cite%
{pak,mle2,mle4,gup6,gup8}. There are many applications of the GUP to physics
such as compact stars, Newtonian gravity, inflationary cosmology, violation
of Lorentz invariance and measurable maximum energy and minimum time
interval \cite{gup3,gup4,gup5,gup11,gup22,gup33,gup44,gup55,gup66}.
Furthermore, after Dvali and Gomez \cite{dvali1,dvali2,dvali3} proposed the
idea of quantum black holes as modeled in Bose-Einstein condensation (BEC)
of marginally bound, self-interacting gravitons, recently one shows that
quantization of gravity is possible by using the Horizon Wave Function (HWF)
formalism \cite{cas1}. Halilsoy et. al. introduced a new metric of
Schwarzschild black hole which is coupled to an external, stationary
electrostatic field by using the interpolation of two exact well-known
solutions of Einstein's equations such as the Schwarzschild (S) metric and a
uniform electromagnetic (em) field solution of Bertotti and Robinson (BR) 
\cite{halilsoy,halilsoy2}. We will here address the Hawking radiation of
entangled particles as an emission of quanta by using this metric.
Furthermore, we investigate the tunneling effect of entangled particles from
such a black hole with the effect of GUP.

Entanglement which plays a frontier role on quantum information, is an
important resource for different computational tasks such as quantum
communication and teleportation. By understanding the entanglement in the
frame of black holes, will give us an important resolution of information
paradox of black holes. On this regard, we propose that two observers, Alice
and Bob, share a generically entangled state at the same initial point.
Behind the coincidence point where the particles tunnel, one of them tunnels
from event horizon of black hole, while the other one stays in the extremal
black hole (BR space-time). We focus our attention on the result of one of the entangled
state which is also equal to another one which may be lost in singularity of
black hole or can tunnel through another universe by a wormhole. The
semi-classical Hawking temperature is derived by applying the WKB
approximation and the Hamilton-Jacobi method to solve the Klein-Gordon (K-G)
and Dirac Equations for the entangled states.

The structure of this paper is as follows. In section II, we briefly give
the black hole solution which will be phrased hereafter as the
Schwarzschild-Electromagnetic black hole, and in section III, by using the
Hamilton-Jacobi method and suitable entangled ansatz, we derive the
corrected Hawking temperature of entangled scalar particles from the new
black hole. Last but not least, in section IV we compute the tunneling rate
of entangled fermion particles \ from the same black hole. Finally, we will
conclude with some comments in section V.

\section{ Schwarzschild-Electromagnetic Black Hole (SEBH)}

The metric for the SEBH in an external electrostatic field in four
dimensions is presented by Halilsoy et. al. as \cite{halilsoy,halilsoy2} 
\footnote{%
The erroneous metric in eq. (15) of ref.(\cite{halilsoy}) is corrected here.}

\begin{equation}
ds^{2}=-fdt^{2}+\frac{1}{f}dr^{2}+r^{2}(d\theta ^{2}+\sin ^{2}\theta
d\varphi ^{2})  \label{mcbh}
\end{equation}%
where 
\begin{equation}
f=1-\frac{2M}{r}+\frac{M^{2}(1-a^{2})}{r^{2}},
\end{equation}

with the external parameter a ($0<a\leq 1$), where the mass M is coupled to
an external em-field. \ Note that $a=0$ is the extremal Reissner--Nordstrom
(RN) case which is transformable to the BR\ metric. The horizon is located
for the above metric at 
\begin{equation}
r_{h}=M(1+a).
\end{equation}

Let us note that the radial coordinate $r$ is related to the vacuum ( say $%
\tilde{r}$ ) coordinate by $r=a\tilde{r}+M(1-a)$. in which $a=0$ is excluded.
Clearly $r$ and $\tilde{r}$ are related by a translation and scaling
transformation. Since $a\leq 1$ the horizon radius is $r_{h}\leq 2m$, which
implies that the em field shrinks the horizon of the Schwarzschild black
hole.

Ricci components are 
\begin{eqnarray}
R_{t}^{t} &=&R_{r}^{r}=\frac{M^{2}(a^{2}-1)}{r^{4}}, \\
R_{\theta }^{\theta } &=&R_{\varphi }^{\varphi }=\frac{-M^{2}(a^{2}-1)}{r^{4}%
}.
\end{eqnarray}%
The Kretschmann scalar which is a quadratic scalar invariant is calculated
as 
\begin{equation}
K=R_{abcd}R^{abcd}=\frac{56M^{2}\left[ (a^{2}-1)^{2}M^{2}+\frac{12}{7}%
Mr\left( a^{2}-1\right) +\frac{6r^{2}}{7}\right] }{r^{8}}
\end{equation}%
The Ricci scalar of the SEBH is calculated as zero (since the
source is pure)%
\begin{equation}
R=0.
\end{equation}

Hence, it is clear that the Einstein tensors are equal to Ricci tensors ( $%
G_{\mu \nu }=R_{_{\mu \nu }}-\frac{1}{2}Rg_{_{\mu \nu }})$ as given\ 
\begin{eqnarray}
G_{t}^{t} &=&G_{r}^{r}=\frac{M^{2}(a^{2}-1)}{r^{4}}, \\
G_{\theta }^{\theta } &=&G_{\varphi }^{\varphi }=\frac{-M^{2}(a^{2}-1)}{r^{4}%
}.
\end{eqnarray}

The following energy-momentum tensor according to Einstein's field equations
\ can be easily obtained by using Einstein tensors ($G_{\mu \nu }$ $=8\pi
T_{_{\mu \nu }})$. Furthermore, the action is 
\begin{equation}
S=\int d^{4}x\sqrt{-g}L
\end{equation}%
where \ $L=\frac{R}{16\pi }-\frac{1}{4}F_{\mu \nu }F^{\mu \nu },$ and \
constants of G and c are 1. The energy-momentum tensors for the vector
potential 
\begin{equation}
A_{\mu }=(\pm \frac{1}{2}\frac{M\sqrt{1-a^{2}}}{ar},0,0,0),
\end{equation}

are defined as 
\begin{equation}
T_{\mu \nu }=F_{\mu \alpha }F_{\nu }^{\alpha }-\frac{1}{4}g_{\mu \nu
}F_{\alpha \beta }F^{\alpha \beta }.
\end{equation}

The corresponding Hawking temperature is found as 
\begin{equation}
T_{H}=\left( \frac{-g_{tt}^{\prime }}{4\pi \sqrt{-g_{tt}g_{rr}}}\right)
_{r=r_{h}}=\frac{1}{4\pi }\frac{2a}{M(a+1)^{2}}.
\end{equation}%
One easily observes that for $a=1$ \ we recover the Schwarzschild result.
For $a=0$ which we have already excluded we obtain $T_{H}=0$, which is
analogous to the extremal RN geometry.

\section{ Entangled Scalar Particles Tunneling from SEBH with GUP}

The modified commutation relation

\begin{equation}
\lbrack x_{i},p_{j}]=i\hbar (1+\alpha p^{2})\delta _{ij}
\end{equation}

is used to derive GUP \cite{kg1,kg2,gup8,kg3} which is given by 
\begin{equation}
\Delta x\Delta p\geq \frac{\hbar }{2}\left\{ 1+\alpha (\Delta p)^{2}\right\}
,
\end{equation}%
where $\alpha =\alpha _{0}/(m_{p}^{2})=\alpha _{0}l_{p}^{2}/\hbar ^{2}$ is a
small value, $m_{p}$ is the Planck mass, $l_{p}$ is the Planck length ($\sim
10^{-35}m$) and $\alpha _{0}<10^{34}$ is a dimensionless parameter.

By using the effect of quantum gravity, we define the generalized
commutation relation to modify KG equation for scalar particles, so to
account for the effects from quantum gravity. The position, momentum, \
energy and frequency operators are \ modified respectively as \cite{kg2,gup8}

\begin{equation}
x_{i}=x_{oi},
\end{equation}

\begin{equation}
p_{i}=p_{0i}(1+\alpha p^{2}),
\end{equation}

\begin{equation}
\varepsilon =E(1+\alpha ^{2}E^{2}),
\end{equation}

and 
\begin{equation}
\bar{\omega}=E(1-\alpha E^{2}),  \label{eq4}
\end{equation}

with the energy operator $E=i\hbar \partial _{0}$.

The square of momentum operators up to order $\alpha ^{2}$ is calculated by 
\begin{eqnarray}
p^{2} &=&-\hbar ^{2}[1-\alpha ^{2}\hbar ^{2}\partial _{j}\partial
^{j}]\partial _{i}[1-\alpha ^{2}\hbar ^{2}\partial _{j}\partial
^{j}]\partial ^{i}  \notag \\
&=&-\hbar ^{2}[\partial _{i}\partial ^{i}-2\alpha ^{2}\hbar ^{2}(\partial
_{j}\partial ^{j})(\partial _{k}\partial ^{k})]+(\alpha ^{4}).  \label{sq}
\end{eqnarray}

where in the last step, we only keep the leading order term of $\alpha $.

Using the generalized Klein-Gordon equation for scalar field in Planck
scale, the generalized K-G equation under the effect of minimum length
having the wave function $\Psi $ can be written as

\begin{equation}
-(i\hslash )^{2}\partial ^{t}\partial _{t}\Psi =\left[ (i\hslash
)^{2}\partial ^{i}\partial _{i}+m_{p}^{2}\right] \left[ 1-2\alpha \left(
(i\hslash )^{2}\partial ^{i}\partial _{i}+m_{p}^{2}\right) \right] \Psi .
\label{14n}
\end{equation}

Herein, using the SEBH metric eq.(\ref{mcbh} )as a background we determine
the entangled scalar particle motion.
An important point in this paper is the entanglement between the infalling and outgoing Hawking particles. The difficulty of this entanglement is to observe inner and outer sides of the black hole horizon. Entangled states can be described by $\Psi$.
After substituting the chosen entangled ansatz of $\Psi $ \cite{ent1,ent2,ent3,ent33}, 
\begin{equation}
\Psi =\kappa e^{\frac{i}{\hbar }S_{A}(t,r,\theta ,\varphi )}+\sqrt{1-\kappa
^{2}}e^{\frac{i}{\hbar }S_{B}(t,r,\theta ,\varphi )}  \label{16n}
\end{equation}

where $S_{A}$ is for Alice and $S_{B}$ is for Bob. Note that $\kappa $ is
some real number which satisfies $|\kappa |\in (0,1)$, so that $\kappa $ and 
$\sqrt{1-\kappa ^{2}}$ are normalized partners. The fate of the Alice and
Bob depends on the value of $\kappa $. The discussion of Alice state is the
same as that of the Bob state. Here, the entanglement can then be observed with only one 
observation for each member of the ensemble so that we just consider the Alice case. To illustrate the idea, we consider the Alice and Bob shares entangled state at the same point in the BR spacetime. Whenever, Alice falls in toward a SEBH and locate near the event horizon of SEBH, as noted here particle state is unentangled when the Hawking temperature is zero and approaches a maximally entangled Bell state as known the black hole evaporates completely \cite{ent4,ent5}. We investigate the fate of this particle by choosing the specific case of $\kappa =1$, which provides us to calculate by using the Alice of entangled ansatz of $\Psi $ inside the K-G solutions
for the SEBH, which is obtained by \cite{kg3} as follows

\begin{eqnarray}
&&\frac{1}{f}\left( \partial _{t}S_{A}\right) ^{2}=f(\partial _{r}S_{A})^{2}+%
\frac{1}{r^{2}}(\partial _{\theta }S_{A})^{2}+\frac{1}{r^{2}\sin ^{2}\theta }%
(\partial _{\varphi }S_{A})^{2}+m_{p}^{2}  \label{kg} \\
&&\times \left[ 1-2\alpha \left( f(\partial _{r}S_{A})^{2}+\frac{1}{r^{2}}%
(\partial _{\theta }S_{A})^{2}+\frac{1}{r^{2}\sin ^{2}\theta }(\partial
_{\varphi }S_{A})^{2}+m_{p}^{2}\right) \right]  \notag
\end{eqnarray}

After it is expanded into the lowest order of $\hbar $ to find the solution
of K-G equation we use the method of separation of variables as follows: 
\begin{equation}
S_{A}(t,r,\theta ,\varphi )=-Et+W(r)+j(\theta ,\phi )+C,  \label{alice}
\end{equation}%
where $C,$ $E$ and $j$ are the \ complex constant, energy and angular
momentum of the scalar particles, respectively.

After substituting eq. (\ref{alice}) into eq. (\ref{kg}), taking only the
radial part yields 
\begin{equation}
\frac{1}{f}E^{2}=f\left( \partial _{r}W\right) ^{2}+m_{p}^{2}\times \left[
1-2\alpha \left( f(\partial _{r}W)^{2}+m_{p}^{2}\right) \right]
\end{equation}
and solving for the $W(r)$, it is found that

\begin{equation}
W_{\pm }=\pm \int dr\frac{1}{f}\frac{\sqrt{E^{2}-m_{p}^{2}(1-2\alpha
m_{p}^{2})f}}{\sqrt{1-2\alpha m_{p}^{2}}}.
\end{equation}

The positive$\textquotedblright +\textquotedblright\ $ signature is for
outgoing entangled Alice scalar particles and the solution with negative

$\textquotedblright -\textquotedblright \ $ signature for the ingoing Alice
scalar particles. Calculating the above integral around the pole at the
horizon by expanding the metric function $f$ about $r_{h}$ ; $f(r_{h})\approx f'(r_{h})(r-r_{h})$ , where prime " ' " denotes a derivative respect to r, gives 
\begin{equation}
W_{\pm }=\pm \frac{i\pi EM(1+a)^{2}}{2a\sqrt{1-2\alpha m_{p}^{2}}}
\end{equation}

While computing the imaginary part of the action, we note that it is same
for both the incoming and outgoing solutions. Herein, a factor two problem
is arisen when calculating the tunneling rate \cite{prama}, however this
problem can be resolved by different method found by Akhmedova et al. \cite%
{singleton1,singleton2}. In this paper we use the most common solution that
if we set the probability of ingoing particles to 100\% (i.e., $P_{-}$ $%
\simeq e^{-2ImW_{-}}= 1$)

Thus, the ingoing and outgoing imaginary action solution is derived as 
\begin{equation}
ImS_{-}=ImW_{-}+ImC=0,
\end{equation}%
and%
\begin{equation}
ImS_{+}=ImW_{+}+ImC
\end{equation}

which result in $ImC=-ImW_{-}$. Contrariwise $W_{+}=-W_{-}$ in order that the
probabilities of radiating particles are obtained as

\begin{equation}
P_{+}=e^{-2ImS}\simeq e^{-4ImW_{+/-}}.
\end{equation}

Now, the probability of particles tunneling from inside to outside the
horizon is given by

\begin{eqnarray}
\Gamma &=&\frac{P_{+}}{P_{-}}\simeq e^{(-4ImW_{+})}  \label{trate} \\
\Gamma &=&e^{-\frac{4i\pi EM(1+a)^{2}}{2a\sqrt{1-2\alpha m_{p}^{2}}}} \label{trate1}
\end{eqnarray}%
and the Hawking temperature for the scalar particles with the effect of
minimum length is obtained as

\begin{equation}
T_{H}=\frac{2a\sqrt{1-2\alpha m_{p}^{2}}}{4\pi EM(1+a)^{2}}.  \label{htem}
\end{equation}

Since ($0<a\leq 1$) it is observed that the electromagnetic field increases
both the tunneling rate (eq. \ref{trate1}) and the Hawking temperature (eq.%
\ref{htem}). When the minimum length effect parameter $\alpha =0,$ it is
equal to the original result of Hawking Temperature. The effect of the
external parameter is shown in figure 1.

\begin{figure}[hb]
\centering
\includegraphics[width=0.80\textwidth]{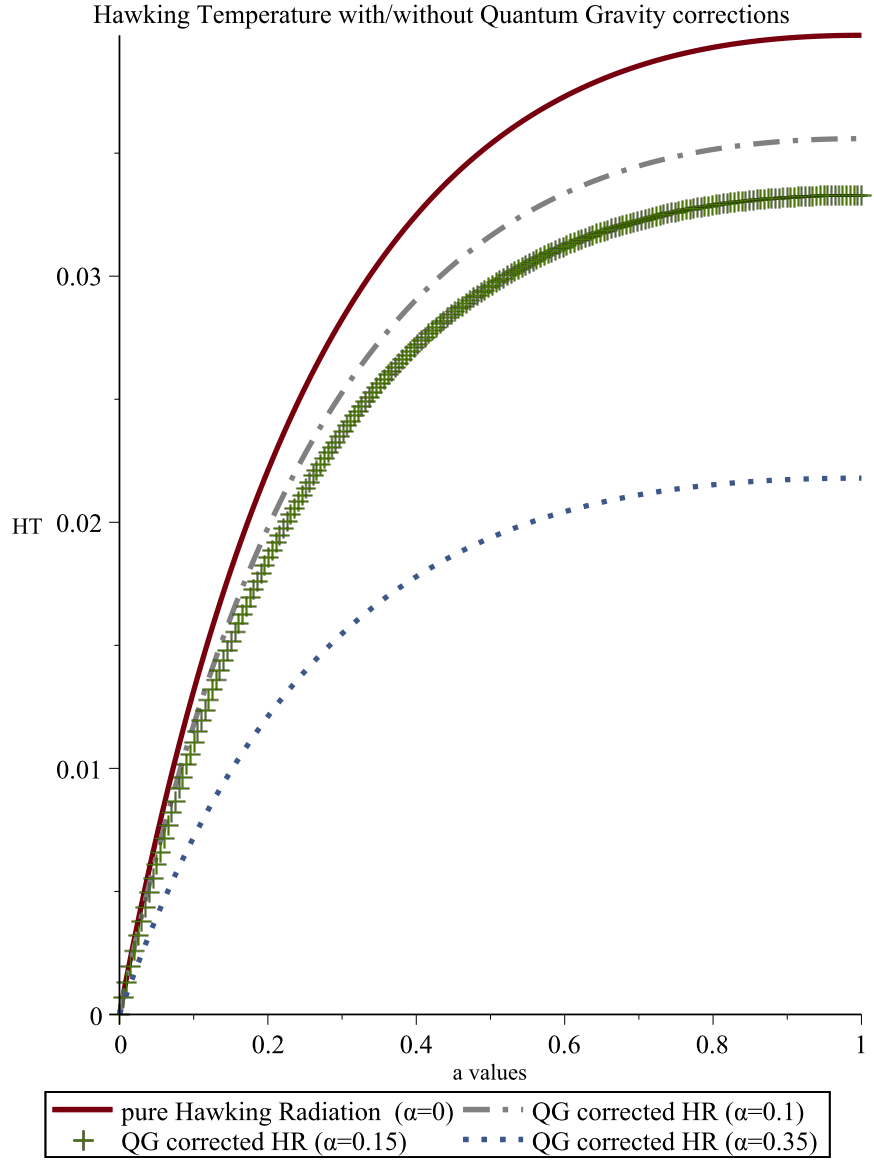}
\caption{The variation of Hawking Temperature against the constant a for $M=E=m_{p}=1$}
\end{figure}

\section{Entangled Fermion Particles Tunneling from SEBH with GUP}

This section makes use of the generalized Dirac equation for fermions \cite%
{gup66,dirac1,dirac2,dirac3} on the background of SEBH \cite{halilsoy}. The
generalized Dirac equation under the effect of minimum length can be written
as

\begin{equation}
-i\gamma ^{t}\partial _{t}\psi =(i\gamma ^{i}\partial _{i}-\frac{1}{2}\gamma
^{\mu }g^{\alpha \beta }\Gamma _{\mu \nu }^{\alpha }J_{\alpha \alpha }+\frac{%
m_{p}}{\hbar })\left[ 1+\alpha \left( \hslash ^{2}\partial ^{i}\partial
_{i}-m_{p}^{2}\right) \right] \psi \text{ ;}i=r,\theta ,\varphi
\end{equation}

where the $\gamma $ matrices are expressed in terms of the Pauli matrices $%
\sigma ^{i}$ as follows 
\begin{eqnarray}
\gamma ^{t} &=&\frac{1}{\sqrt{f}}\left( 
\begin{array}{cc}
i & 0 \\ 
0 & -i%
\end{array}%
\right) ,\,\,\ \gamma ^{r}=\sqrt{g}\left( 
\begin{array}{cc}
0 & \sigma ^{3} \\ 
\sigma ^{3} & 0%
\end{array}%
\right) ,  \notag \\
\gamma ^{\theta } &=&\frac{1}{r}\left( 
\begin{array}{cc}
0 & \sigma ^{1} \\ 
\sigma ^{1} & 0%
\end{array}%
\right) ,\,\,\,\ \gamma ^{\varphi }=\frac{1}{r\text{sin}\theta }\left( 
\begin{array}{cc}
0 & \sigma ^{2} \\ 
\sigma ^{2} & 0%
\end{array}%
\right) ,  \label{5n}
\end{eqnarray}

and \ 
\begin{equation}
J_{\alpha \alpha }=\frac{i}{4}\left[ \gamma ^{\alpha },\gamma ^{\beta }%
\right] ,\{\gamma ^{\mu },\gamma ^{\nu }\}=2g^{\mu \nu },
\end{equation}

\begin{equation}
\Gamma _{tt}^{r}=\frac{f^{\prime }f}{2},\Gamma _{tr}^{t}=\frac{f^{\prime }}{%
2f}.
\end{equation}

Consequently, the generalized Dirac equation can be written as \cite{dirac3}

\begin{multline}
i\gamma ^{t}\partial _{t}+i\gamma ^{i}\partial _{i}\left( 1-\alpha
m_{p}^{2}\right) +i\gamma ^{i}\alpha \hbar ^{2}\left( \partial _{j}\partial
^{j}\right) \partial _{i} \\ 
 +\frac{m_{p}}{\hbar }\left( 1+\alpha \hbar
^{2}\partial _{j}\partial ^{j}-\alpha m_{p}^{2}\right) 
+i\gamma ^{\mu }\Omega _{\mu }\left( 1+\alpha \hbar ^{2}\partial
_{j}\partial ^{j}-\alpha m_{p}^{2}\right) \psi=0.
\end{multline}

By using the ansatz for the entangled spin-up $\psi $ and only the r
direction, one obtains

\begin{equation}
\psi _{\uparrow }=\kappa \left( 
\begin{array}{c}
c_{0} \\ 
\\ 
c_{2} \\ 
\end{array}%
\right) e^{\frac{i}{\hbar }S_{A}(t,r)}+\sqrt{1-\kappa ^{2}}\left( 
\begin{array}{c}
c_{4} \\ 
\\ 
c_{6} \\ 
\end{array}%
\right) e^{\frac{i}{\hbar }S_{_{B}}(t,r,\theta ,\varphi )},  \label{7n}
\end{equation}

with constant spinor components $c_{i}$, $i=0,2,4,6$.

Upon choosing the specific case of $\kappa =1$ ( the Alice case) of
entangled ansatz of $\Psi $ \ and solve the generalized Dirac equation on
the background of SEBH, taking the lowest order of $\hbar $ , it is obtained
that the set of equations \ for the form depending only on the radial part
of $S_{A}(t,r)=Et+W(r)$ \ are \cite{kg3}

\begin{multline}
\alpha ^{2}f^{4}(\partial _{r}W)^{6}+\alpha f^{3}(3m_{p}^{2}\alpha
-2)(\partial _{r}W)^{4}  
\\
+f^{2}[(1-\alpha m_{p}^{2})^{2}-\alpha (2m_{p}^{2}-2m_{p}^{4}\alpha
)](\partial _{r}W)^{2}+m_{p}^{2}f(1-\alpha m_{p}^{2})^{2}-E^{2} =0
\end{multline}

To analyze the solution of $W(r)$, \ by neglecting the higher order terms of 
$\alpha $, the solution for $W(r)$ \ is calculated as 
\begin{equation}
W_{\pm }=\pm \int dr\frac{1}{f}\frac{\sqrt{E^{2}-m_{p}^{2}(1-2\alpha
m_{p}^{2})f}}{\sqrt{1-2\alpha m_{p}^{2}}}.
\end{equation}%
\begin{equation}
W_{\pm }=\pm \frac{i\pi EM(1+a)^{2}}{2a\sqrt{1-2\alpha m_{p}^{2}}}
\end{equation}

Now, using the same procedure in eq.(28-30), the probability of the particles going out of horizon from
inside is written as

\begin{eqnarray}
\Gamma &=&\frac{P_{+}}{P_{-}}\simeq e^{(-4ImW_{+})} \\
\Gamma &=&e^{-\frac{4i\pi EM(1+a)^{2}}{2a\sqrt{1-2\alpha m_{p}^{2}}}}
\end{eqnarray}

and the corresponding Hawking temperature is 
\begin{equation}
T_{H}=\frac{2a\sqrt{1-2\alpha m_{p}^{2}}}{4\pi EM(1+a)^{2}}.
\end{equation}

\section{Conclusion}

In summary, by using the modified Klein-Gordon and Dirac equations under the
effect of quantum gravity, we have examined the entangled scalar/fermion
particle's tunneling from SEBH. The generalized uncertainty principle and
application on the fields are used to derive corrected Hawking radiation
with the help of Hamilton-Jacobi method. Entangled particles such as Alice
and Bob particles can tunnel from the black hole with an equivalent energy.
Charge, mass and energy of the tunneled entangled particles are only
properties. Furthermore, it is easy to conclude that during the evaporation,
temperature increase is decelerated by the effect of GUP. Hence, it is
understood that the two effects will be canceled at some point in the
radiation and remnants are left. In addition, the external parameter of SEBH
"a" can be chosen as the value of ($0 < a\leq 1$) to change the Hawking
temperature. This is how the Hawking temperature modifies when the
Schwarzschild black hole is immersed in a uniform external electromagnetic
field specified by the parameter "a". All our results reduce to that of
Schwarzschild case whenever we set $a=1$.

\section{Acknowledgments}

I am indebted to my supervisor M. Halilsoy for continuous suggestions and
encouragements. Valuable discussions with Gerard 't Hooft, G. Dvali, George
F. R. Ellis, R. Mann, E. Guendelman, R. Casadio, O. Panella, I. Sakalli, A.
Addazi and A. Helou are gratefully acknowledged. The research was started
while the author was at the Karl Schwarzschild Meeting 2015 (KSM2015) at
FIAS in Germany. I also would like to thank organizers of KSM2015 in
Frankfurt for their hospitality.

I am thankful also to an anonymous referees whose comments helped to improve
the paper.

\end{document}